\documentclass[runningheads]{llncs}
\usepackage{amssymb}
\usepackage{amsmath}
\usepackage{graphicx}
\usepackage{epstopdf}
\usepackage{multirow}
\begin{document}

\title{BEFD: Boundary Enhancement and Feature Denoising for Vessel Segmentation}
\titlerunning{BEFD: Boundary Enhancement and Feature Denoising}
\author{Mo Zhang$^{1,2,3}$, Fei Yu$^{1}$, Jie Zhao$^{2}$, Li Zhang$^{2\dagger}$, Quanzheng Li$^{4\dagger}$}
\institute{$^{1}$Center for Data Science, Peking University, Beijing 100871, China; $^{2}$Center for Data Science in Health and Medicine, Peking University, Beijing 100871, China; $^{3}$Laboratory for Biomedical Image Analysis,Beijing Institute of Big Data Research, Beijing 100871,  China; $^{4}$Center for Advanced Medical Computing and Analysis, MGH/BWH Center for Clinical Data Science, Department of Radiology, Massachusetts General Hospital, Harvard Medical School, Boston, MA 02115, USA.\\
$\dagger$ Joint corresponding authors \\
\email{zhangli\_pku@pku.edu.cn}\\
\email{li.quanzheng@mgh.harvard.edu}}
\authorrunning{M. Zhang et al.}
\maketitle
\begin{abstract}
Blood vessel segmentation is crucial for many diagnostic and research applications. In recent years, CNN-based models have leaded to breakthroughs in the task of segmentation, however, such methods usually lose high-frequency information like object boundaries and subtle structures, which are vital to vessel segmentation. To tackle this issue, we propose Boundary Enhancement and Feature Denoising (BEFD) module to facilitate the network ability of extracting boundary information in semantic segmentation, which can be integrated into arbitrary encoder-decoder architecture in an end-to-end way. By introducing Sobel edge detector, the network is able to acquire additional edge prior, thus enhancing boundary in an unsupervised manner for medical image segmentation. In addition, we also utilize a denoising block to reduce the noise hidden in the low-level features. Experimental results on retinal vessel dataset and angiocarpy dataset demonstrate the superior performance of the new BEFD module.
\end{abstract}
\section{Introduction}
 The precise segmentation of blood vessel plays an important role in the diagnosis of related diseases. For example, the morphological changes of retinal vessel may indicate some relevant diseases (eg. glaucoma, diabetes and hypertension), and the  quantitative analysis of coronary arteries in digital subtraction angiography (DSA) is commonly used in the assessment of myocardial infarction and coronary atherosclerotic disease. Moreover, some pathophysiological changes of retinal vessel caused by prolonged hyperglycemia are related to the smallest vessels \cite{archer1999diabetic}, so it is vital to extract the thin vessels. In the past decades, many automatic approaches have been proposed to segment the blood vessel, such as tracking based models \cite{chutatape1998retinal}, filtering based models \cite{chaudhuri1989detection} and deformable models \cite{dizdaro2012level}. 
 
 Recently, deep learning based methods have significantly improved the performance of vessel segmentation. For example, Hu \textit{et al.} \cite{hu2018retinal} presented a multi-scale convolutional neural network (CNN) with an improved cross-entropy loss for vessel segmentation, outperforming the traditional unsupervised algorithms. However, although CNN has tremendous power to extract high-level representative features, it is inevitable that spatial information is partly lost in the downsampling layers. Further, the lost information mainly belongs to the high-frequency components in the image, such as object boundary which contains essential cues for localization. This is consistent with the observation in \cite{zhang2018deep}, where they found that most mispredicted pixels by UNet are located in the edges of blood vessels. Therefore, it is very necessary to strengthen the network capability to capture edge information in vessel segmentation.
 
There have been presented a variety of related methods, which can be roughly divided into four categories: 1) Many works reused the low-level features with rich spatial information by a ``skip connection" structure to maintain more edge details, such as SegNet \cite{badrinarayanan2017segnet}, UNet \cite{ronneberger2015u} and DeeplabV3+ \cite{chen2018encoder}; 2) Some approaches employed post-processing techniques (eg. conditional random fields (CRFs) \cite{chen2017deeplab}, active contour models \cite{sun2018extracting}) to refine the segmentation results; 3) Some researchers exploited an additional branch for edge detection to enhance the segmentation task, which can be regarded as a multi-task framework \cite{hatamizadeh2019end,qin2019transfer,zhu2019boundary}; 4) Some studies proposed a boundary-aware loss function \cite{zhen2019learning} or labeled the object contours as an independent class \cite{zhang2018deep}, thus paying more attention to the border pixels.  

In this paper, we provide a novel solution to boost the accurate delineation of borders in blood vessel segmentation, which serves as an insertable module called Boundary Enhancement and Feature Denoising (BEFD). By applying the traditional edge detection operator to the raw image, BEFD can gain an edge attention map in an unsupervised way. This pixel-wise attention map assigns higher weights to the pixels around the object boundary, which would be multiplied (element-wise product) by certain intermediate feature maps in the encoder part. In this way, the prior knowledge about vascular edges are transferred to the neural networks, enhancing the boundary information in semantic segmentation. In addition, in order to avoid amplifying noise at the same time, BEFD also conducts feature denoising in the process of skip connection via a denoising block. 

We evaluate the proposed BEFD module on both retinal vessel images and angiocarpy images. The experimental results demonstrate the effectiveness of the BEFD module. Compared to existing state-of-the-art models for vessel segmentation, UNet integrated with BEFD module achieves the highest score in accuracy, AUC, sensitivity and specificity on DRIVE dataset.

The contributions of this work are summarized as follows:\\
1) We propose an innovative module (BEFD) to boost boundary feature extraction in semantic segmentation, which can be easily incorporated into any encoder-decoder framework in an end-to-end manner. Moreover, the proposed BEFD module applies the unsupervised edge detector to provide edge prior knowledge for CNNs in medical image segmentation.\\ 
2) We use the denoising block to eliminate the noise existing in the low-level feature maps, preserving the indeed required spatial information for segmentation.\\
3) Our approach obtains remarkable performance for blood vessel segmentation, actually, it can also be easily generalized to deal with other segmentation tasks.  
\section{Method}
\begin{figure}[t]
	\centering
	\includegraphics[width =\textwidth]{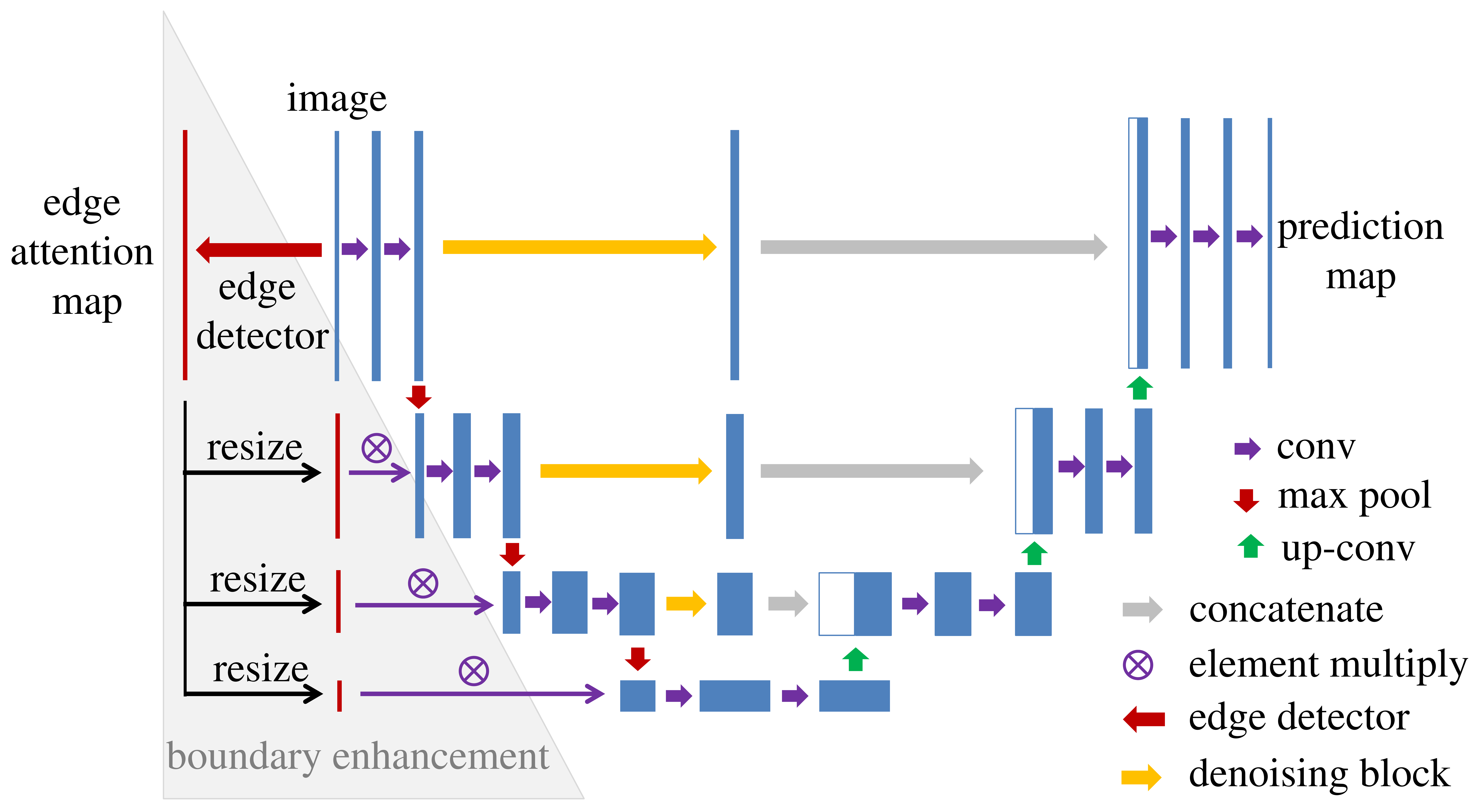}
	\caption{Architecture of BEFD-UNet. It consists of three parts: 1) the basic UNet; 2) the boundary enhancement (BE) part in the gray triangle, which employs an edge detector to provide boundary localization for the encoder path of UNet; 3) the feature denoising (FD) part, formed by three denoising blocks in the phase of skip connection.}
	\label{fig:network}
\end{figure}
\subsection{The architecture of BEFD-UNet}
In this section, we choose U-Net as baseline, to elaborate how the BEFD module can fit into it. The incorporated model is named as BEFD-UNet, and its structure is illustrated in Fig.~\ref{fig:network}.  The baseline UNet consists of an encoder path and a decoder path. In the encoder path, each step includes two $3 \times 3$ convolutions followed by a $2 \times 2$ max pooling with stride of 2 for downsampling, which doubles the number of feature map channels. In the decoder path, each layer contains a deconvolution for upsampling followed by two $3 \times 3$ convolutions, where the channel number is halved. Additionally, the skip connection between encoder and decoder combines the low-level features in the shallow layers with the high-level abstract features, preserving more spatial information for better localization.

As for the BEFD module, it is composed of two parts: the Boundary Enhancement (BE) part and Feature Denoising (FD) part. In the BE part, edge detection is firstly conducted on the raw image to get a pixel-wise edge attention map, whose values indicate the importance of the corresponding pixels, in this setting, edge pixels are assigned with greater weights. Then, this attention map is incorporated into the last three layers in the encoder path via element-multiplication, before which the attention map has been resized to be consistent with the corresponding feature maps by bilinear interpolation. Benefiting from the edge prior produced by the unsupervised edge detection, the boundary information is enhanced with little extra computation. 

In the FD part, the skip connection structure is modified by adding a denoising block before the simple concatenation. The low-level features in the encoder path contain not only rich spatial details, but also background noise which is undesirable for the decoder path, so it is necessary to make feature denoising in the process of skip connection. On the other hand, edge detector may misidentify image noise as boundary, thus enhancing the unexpected noise simultaneously, in this view, feature denoising is also indispensable. Furthermore, the denoising block \cite{xie2019feature} is a typical method to restrain the noise in the intermediate feature maps, so we apply it to the low-level features. In addition, more details about the edge detection and feature denoising are described in the next sections.   
\subsection{Unsupervised edge detection}
Edge detection aims to locate the object boundaries in the image, which is an important step towards extracting image features. In this paper, we utilize Sobel edge detector \cite{kittler1983accuracy} to obtain this goal. Sobel edge detector calculates the first derivatives of the raw image for the horizontal direction and vertical direction separately, then combines these two components together via the absolute magnitude of gradient. It can be expressed as follow:
\begin{equation}
G_{x}=\left[\begin{array}{lll}
-1 & 0 & 1 \\
-2 & 0 & 2 \\
-1 & 0 & 1
\end{array}\right] * I,
G_{y}=\left[\begin{array}{ccc}
-1 & -2 & -1 \\
0 & 0 & 0 \\
1 & 2 & 1
\end{array}\right] * I,
G=\sqrt{G_{x}^{2}+G_{y}^{2}}
\end{equation}
where $I$ is the raw image, $*$ denotes the convolution operator, $G_x,G_y$ are the gradient components of $x$ axis and $y$ axis respectively, and $G$ denotes the Sobel edge map. In order to obtain the final edge attention map which emphasizes the boundary pixels in a suitable and adjustable way, we apply the thresholding method as well as linear transformation to the Sobel edge map as follows:
\begin{equation}\begin{aligned}
G_{final}(x,y) &=\left\{\begin{array}{ll}
{1,} & {\text { if } G(x,y) > \lambda_{max} \text { or } < \lambda_{min}} \\
{(1-\frac{G(x,y)-\lambda_{min}}{\lambda_{max}-\lambda_{min}})\cdot \alpha + \beta,} & {\text {otherwise}}
\end{array}\right.
\end{aligned}\end{equation}
Here $G(x,y)$ denotes the Sobel gradient value at pixel $(x,y)$, and $\lambda_{min},\lambda_{max},\alpha,\beta$ are tunable parameters to regulate the scope of weights. Note that in the setting of $1-\frac{G(x,y)-\lambda_{min}}{\lambda_{max}-\lambda_{min}}$, weak edges are emphasized with higher weights than the strong ones, as the key challenge in vessel segmentation is the detection of micro vessels with low contrast. 
\subsection{Feature denoising}
\begin{figure}[!t]
	\centering
	\includegraphics[width =0.5\textwidth]{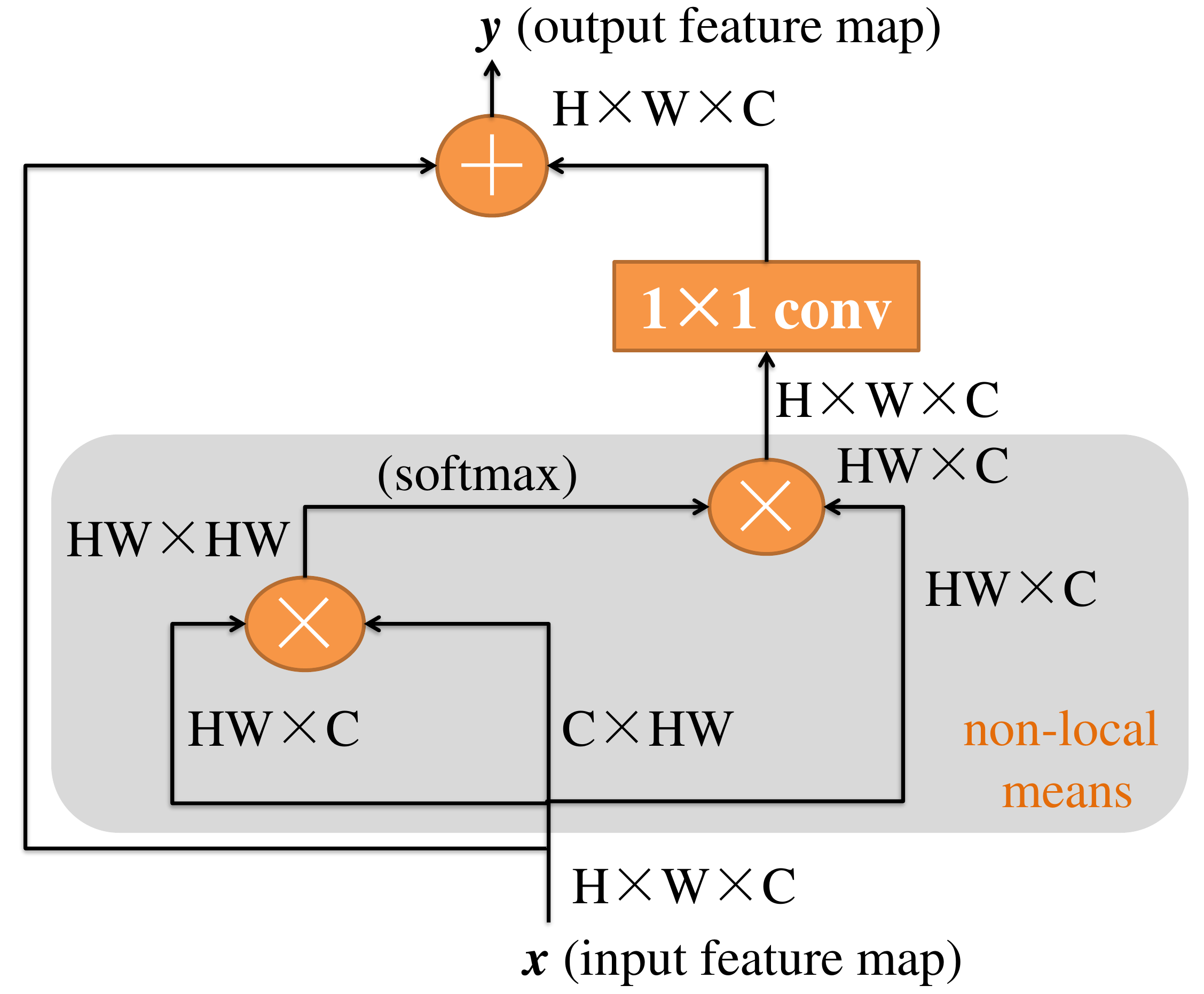}
	\caption{Architecture of the feature denoising block \cite{xie2019feature}.}
	\label{fig:nonlocal}
\end{figure}
The concept of feature denoising \cite{xie2019feature} is proposed to improve adversarial robustness in image classification, based on the observation that adversarial perturbations on images result in noise in the features. At the same time, segmentation networks always make efforts to maintain more high-frequency components for sake of restoring more details (eg. skip connection in UNet \cite{ronneberger2015u}), providing an opportunity for the introduction of noise in the feature maps. Hence, it is also necessary to conduct feature denoising in image segmentation as we do in this work. 

Non-local means \cite{buades2005non} is widely used for the task of image denoising, which calculates a weighted mean of all locations in the image. It is defined as:
\begin{equation}
\mathbf{y}_{i}=\frac{1}{\mathcal{C}(\mathbf{x})} \sum_{\forall j} f\left(\mathbf{x}_{i}, \mathbf{x}_{j}\right) \cdot \mathbf{x}_{j},
\end{equation}
where $\mathbf{y}_{i}$ is the output value at location $i$, $\mathbf{x}_{i}$ is the input value at location $i$, and $j$ denotes all possible positions. $f$ is a weighting function and $\mathcal{C}(\mathbf{x})$ is a normalization factor. Following the idea of non-local means \cite{buades2005non} and non-local neural networks \cite{wang2018non}, the work \cite{xie2019feature} presented a denoising block to make feature denoising. As shown in Fig.2, except for the non-local means part, the denoising block also contains a $1\times 1$ convolution and residual connection for feature fusion. In this work, we use this denoising block to eliminate the feature noise in segmentation networks, where the pairwise function $f$ is set to dot product $f\left(\mathbf{x}_{i},\mathbf{x}_{j}\right)=\mathbf{x}_{i}^{\mathrm{T}} \mathbf{x}_{j}$ and $\mathcal{C}(x)=N$ ($N$ is the number of locations in $\mathbf{x}$). More information about the denoising block can be found in \cite{xie2019feature,wang2018non}.
\section{Experiments}
To evaluate the effectiveness of the BEFD module, we compare two models: 1) the baseline UNet; 2) BEFD-UNet, which integrates BEFD module into UNet.\\  
\\
\textbf{Data Description.} 
We evaluate the new approach on two datasets of blood vessel: DRIVE and HEART. The DRIVE dataset is a public resource including 40 fundus images and corresponding labels of retinal vessel (the first manual annotation). The resolution of each image is $565\times 584$, and the whole set has originally been divided into two parts: 20 images for training and the rest 20 images for testing. Apart from the fundus images, we also collect 1092 digital subtraction angiographies (DSA) of coronary (546 for train, 218 for valid and 328 for test). All images are resized to the same dimension $256\times 256$, and this dataset is named as HEART. Moreover, both datasets are preprocessed by Contrast Limited Adaptive Histogram Equalization (CLAHE) and normalization. To avoid overfitting, data augmentation is used by horizontal and vertical flip.\\
\\
\textbf{Evaluation Metrics.} 
To quantify the performance of our approach, we use several metrics consisting of sensitivity (Sen), specificity (Spe), accuracy (Acc), and F1-score (F1), the formulas are shown below:
\begin{equation}\begin{aligned}
&Acc=\frac{TP+TN}{TP+TN+FP+FN},Sen=\frac{TP}{TP+FN},\\
&Spe=\frac{TN}{TN+FP},F1=\frac{2TP}{2TP+FP+FN},
\end{aligned}\end{equation}
where TP,TN,FP,FN denote the numbers of true positives, true negatives, false positives and false negatives respectively. We also calculate the AUC metric (the area under the ROC curve) for evaluation. Note that these metrics are calculated over the whole image. In addition, the DRIVE dataset provides masks for labels, and the performance evaluated within the masks is shown in Table 1 of the supplementary material. \\
\\
\begin{table}[!b]
\begin{center}
\caption{Quantitative analysis of different methods on DRIVE dataset.}
\begin{tabular*}{0.70\linewidth}{c|c|ccccc}
\hline
Methods & Year & Sen & Spe & F1 & Acc & Auc  \\ 
\hline
Hu\cite{hu2018retinal} & 2018 & 0.7772 & 0.9793 & -  & 0.9533   & 0.9759  \\
Zhuang\cite{zhuang2018laddernet} & 2018 & 0.7856 & 0.9810  & 0.8202 & 0.9561   & 0.9793   \\
Alom\cite{alom2019recurrent} & 2019  & 0.7792 & 0.9813 & 0.8171 & 0.9556   & 0.9784  \\
Jin\cite{jin2019dunet} & 2019 & 0.7963  & 0.9800 & 0.8237 & 0.9566   & 0.9802     \\
Mou\cite{mou2019dense} & 2019 & 0.8126 & 0.9788  & - & 0.9594   & 0.9796   \\
Wu\cite{wu2019vessel} & 2019 & 0.8038 & 0.9802  & - & 0.9578   & 0.9821   \\
Lyu\cite{lyu2019fundus} & 2019 & 0.7940 & 0.9820  & - & 0.9579   & 0.9826   \\
Wang\cite{wang2019dual}& 2019 & 0.7940 & 0.9816  & \textbf{0.8270} & 0.9567   & 0.9772   \\
Zhou\cite{zhou2019symmetric}& 2019 & 0.8135 & 0.9768  & 0.8249 & 0.9560   & 0.9739   \\
Li\cite{li2020iternet}& 2020 & 0.7791 & 0.9831  & 0.8218 & 0.9574   & 0.9813   \\
\hline
UNet & 2020 & 0.7887   & \textbf{0.9861} & 0.8140 & 0.9686 & 0.9836\\
\textbf{BEFD-UNet} & 2020 & \textbf{0.8215} & 0.9845 & 0.8267 & \textbf{0.9701}   & \textbf{0.9867}   \\
\hline
\end{tabular*}
\label{tab:table1}
\end{center}
\end{table}
\begin{table}[!b]
\begin{center}
\caption{Quantitative analysis of different methods on HEART dataset.}
\begin{tabular*}{0.64\linewidth}{c|ccccc}
\hline
Methods & Sen & Spe  & F1 & Acc & Auc  \\ 
\hline
UNet &  0.9186   & 0.9839 & 0.9073 & 0.9750 & 0.9938\\
BE-UNet &  \textbf{0.9411}   & 0.9779 & 0.9030 & 0.9730 & 0.9935\\
FD-UNet &  0.9234   & \textbf{0.9849} & 0.9140 & 0.9767 & 0.9939\\
\textbf{BEFD-UNet} &  0.9333   & 0.9835 & \textbf{0.9141} & \textbf{0.9767} & \textbf{0.9947} \\
\hline
\end{tabular*}
\label{tab:table2}
\end{center}
\end{table}
\begin{figure}[t]
	\centering
	\includegraphics[width =1\textwidth]{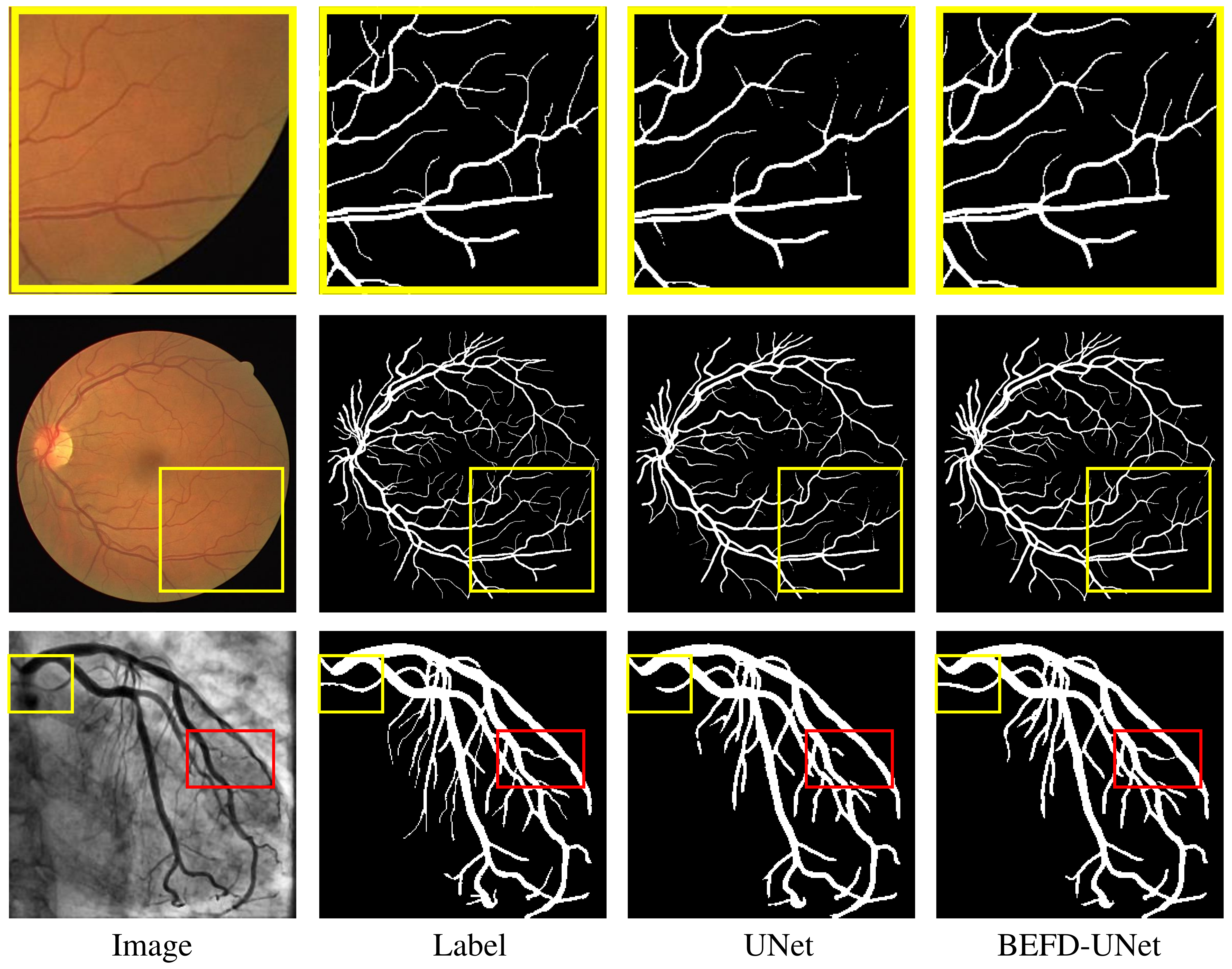}
	\caption{Segmentation results of different models. The top two rows: prediction maps on DRIVE dataset; The bottom row: prediction maps on HEART dataset.}  
	\label{fig:compare}
\end{figure}
\textbf{Implementation Details.} 
All networks are implemented in TensorFlow 1.2.1
(https://www.tensorflow.org) on  a  single  NVIDIA  GTX1080ti GPU. We use the Adam algorithm to minimize the cross entropy loss, which is trained for 30k iterations with batch size of 8. In order to accelerate convergence, batch normalization is followed by each convolutional layers. In the UNet architecture, we set the channel number in the first layer to 64. The parameters $\lambda_{min},\lambda_{max},\alpha,\beta$ in edge detection are set to 0.8, 5.0, 2.0 and 1.0 respectively.
\section{Results}
\textbf{Evaluation of the performance of the BEFD module.} 
Table~\ref{tab:table1},~\ref{tab:table2} report the qualitative results of retinal vessel segmentation on DRIVE dataset and cardiac vessel segmentation on HEART dataset respectively. Compared to the baseline UNet, BEFD-UNet obtains better performance in four of the total five metrics with a considerable margin on both datasets, only Spe (specificity) is a bit lower. Moreover, BEFD-UNet provides a significant improvement in Sen (sensitivity) from 0.7887 (0.9186) to 0.8215 (0.9333) on DRIVE dataset (HEART dataset), indicating that our approach yields a lower false negative (FN) ratio. This can also be demonstrated by the segmentation results shown in Fig.~\ref{fig:compare}, where BEFD-UNet successfully captures the extremely thin retina vessels and identifies the coronary arteries with low contrast to background. Besides the evaluation of the whole image, we also evaluate the models on small vessels (with a width of 1 or 2 pixels) only. In such a setting, BEFD-UNet makes a significant improvement in F1 score from 0.7163 (UNet) to 0.7628 (BEFD-UNet) on DRIVE dataset, indicating that our method has strong ability to capture the thin vessels.   

On the other hand, we also compare the results with recent state-of-the-art models on DRIVE dataset as listed in Table~\ref{tab:table1}. It shows that the proposed BEFD-UNet ranks first in four metrics except for F1 score. More specifically, it achieves the highest accuracy (1.05\% higher than the second best \cite{mou2019dense}), the optimal AUC (0.41\% higher than the suboptimal result \cite{lyu2019fundus}) as well as the best sensitivity (0.80\% higher than the previous highest score \cite{zhou2019symmetric}). Additionally, it is worth mentioning that the F1 score (0.8267) of our method is relatively close to the best score (0.8270) \cite{wang2019dual}. The above comparisons demonstrate the strong capacity of the new BEFD module to tackle semantic segmentation.\\
\\
\textbf{Discussion about the mechanism of the BEFD module.} 
\begin{figure}[!t]
	\centering
	\includegraphics[width =0.8\textwidth]{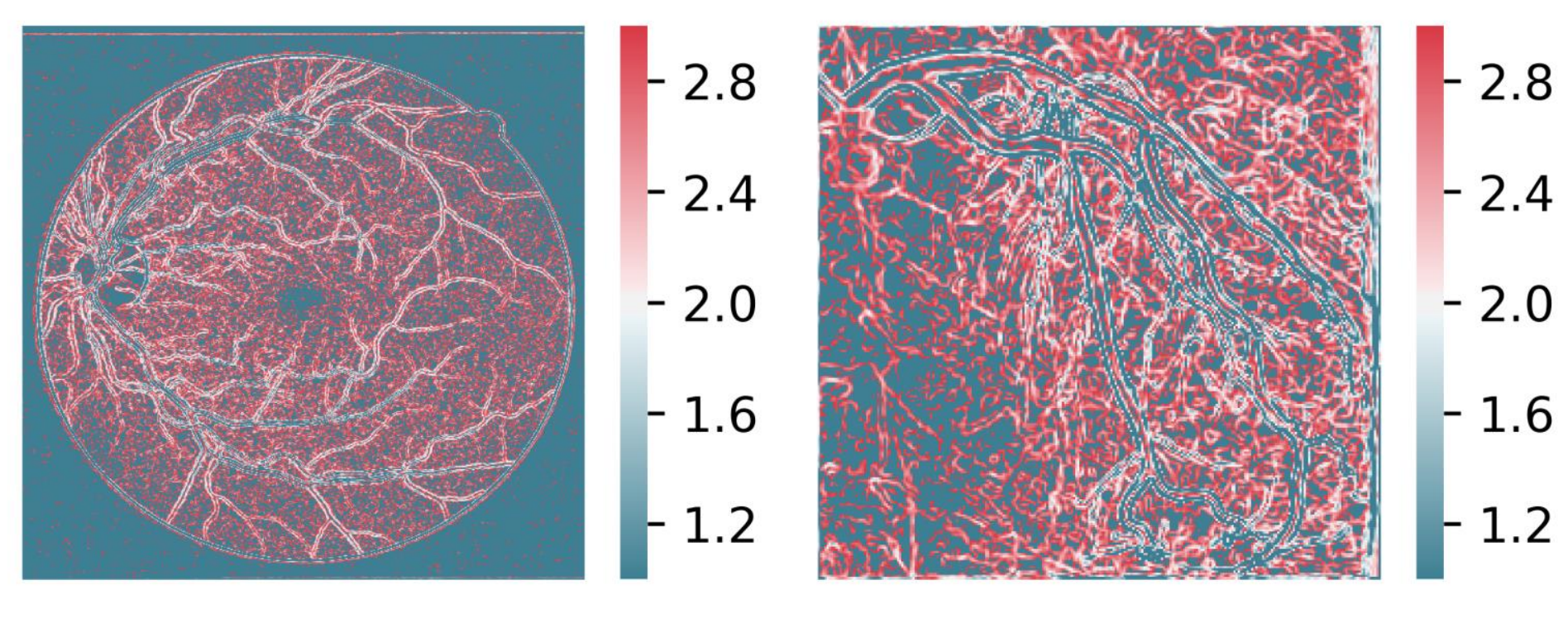}
	\caption{Example of edge attention map on DRIVE dataset (left) and HEART dataset (right). Boundary pixels (mostly white) have higher values than background pixels (blue), while noise (red points) is widely distributed on the map, requiring feature denoising to restrain it.}  
	\label{fig:edgemap}
\end{figure}
To explore the effect of the BEFD module, we show an example of edge attention map on both datasets in Fig.~\ref{fig:edgemap}. It can be observed that boundary pixels have higher values, making it possible to pay more attention to the object contours. When the feature maps in the encoder path are multiplied (element-wise) by such attention map, the object boundaries are enhanced accordingly. At the same time, maps contain lots of noise due to the emphasis on weak vessels, for which it requires introducing the feature denoising block. In addition, we also conduct experiments using the boundary enhancement (BE) part or feature denoising (FD) part independently, which are shown in Table~\ref{tab:table2}. The 
results indicate that using either BE or FD part would produce an unbalanced result. BE-UNet tends to generate high Sensitivity with low Specificity, because the BE part may misidentify image noise as boundaries, thus amplifying the undesirable noise. On the other hand, FD-UNet performs oppositely (high Specificity with low Sensitivity), since the FD block may make excessive denoising, which eliminates some fine structures. Therefore, we apply both BE and FD blocks jointly in BEFD-UNet, which leverages the advantages of both the BE and FD blocks.
\section{Conclusion}
In this paper, we propose a novel BEFD module to boost the boundary localization in the encoder-decoder framework for blood vessel segmentation. The integrated BEFD-UNet outperforms the baseline UNet as well as most of state-of-the-art approaches, resulting from its powerful ability to detect extremely tiny vessels. More broadly, the BEFD module provides a novel solution to leverage the advantage of traditional image processing algorithm, which can compensate for the defects of CNNs in an unsupervised way. This mechanism is worth investigating further in the future work.
\subsubsection*{Acknowledgments.} This work was supported by Natural Science Foundation of China (NSFC) under Grants 81801778, 71704024, 11831002; National Key R\&D Program of China (No. 2018YFC0910700); Beijing Natural Science Foundation (Z180001).
\bibliographystyle{splncs04}
\bibliography{paper1469}

\clearpage
\title{Supplementary Material}
\author{}
\institute{}
\maketitle 
\begin{table}
\begin{center}
\caption{Quantitative analysis of different methods on DRIVE dataset. The performances of models are evaluated inside the masks. }
\begin{tabular*}{0.60\linewidth}{c|ccccc}
\hline
Methods &  Sen & Spe & F1 & Acc & Auc  \\ 
\hline
UNet &  0.7891   & 0.9789 & 0.8143 & 0.9544 & 0.9768\\
BEFD-UNet &  0.8218 & 0.9765 & 0.8269 & 0.9565   & 0.9803   \\
\hline
\end{tabular*}
\label{tab:table3}
\end{center}
\end{table}

\end{document}